\documentclass[12pt]{article}
\usepackage{graphicx}
\usepackage[bookmarksnumbered,colorlinks,plainpages]{hyperref}

\begin{document}

\title{{\small Physica A (2004), in press}\\
Fractal geometry, information growth and nonextensive thermodynamics}

\author{Q.A. Wang, L. Nivanen, A. Le M\'ehaut\'e, \\
{\it Institut Sup\'erieur des Mat\'eriaux et M\'ecaniques Avanc\'es}, \\
{\it 44, Avenue F.A. Bartholdi, 72000 Le Mans, France} \\
and M. Pezeril \\ {\it Laboratoire de Physique de l'\'etat Condens\'e}, \\
{\it Universit\'e du Maine, 72000 Le Mans, France}}

\date{}

\maketitle

\begin{abstract}
This is a study of the information evolution of complex systems by a geometrical
consideration. We look at chaotic systems evolving in fractal phase space. The
entropy change in time due to the fractal geometry is assimilated to the information
growth through the scale refinement. Due to the incompleteness of the state number
counting at any scale on fractal support, the incomplete normalization
$\sum_ip_i^q=1$ is applied throughout the paper, where $q$ is the fractal dimension
divided by the dimension of the smooth Euclidean space in which the fractal
structure of the phase space is embedded. It is shown that the information growth is
nonadditive and is proportional to the trace-form $\sum_ip_i-\sum_ip_i^q$ which can
be connected to several nonadditive entropies. This information growth can be
extremized to give power law distributions for these non-equilibrium systems. It can
also lead to a nonextensive thermodynamics for heterogeneous systems which contain
subsystems each having its own $q$. It is shown that, within this thermodynamics,
the Stefan-Boltzmann law of blackbody radiation can be preserved.
\end{abstract}


\vspace{1cm}

\section{Introduction}
The aim of this work is to study the information evolution and the thermodynamics of
non-equilibrium complex systems whose phase space volume gradually maps into
fractals or multifractals at long time $t\rightarrow\infty$\cite{Hilb94}.

The motivation of this work is relevant to the generalization of Shannon information
entropy which are in general connected to dense phase space ($\Gamma$-space). One
notices that the generalized entropies\cite{Esteban} are just posited or postulated
as such and recover Shannon entropy as the generalization indexes (parameters) take
special values. Some of these generalized entropies, e.g., Havrda-Charvat-Tsallis
one\cite{Havrda,Tsal88} and R\'enyi one\cite{Reny66}, are believed to have some
connections with fractal geometry of phase space and with chaotic behavior of
non-equilibrium systems at stationary states, and have been used to develop
generalized statistics whose mathematical structure mimics the formal system of the
conventional statistical mechanics of Boltzmann-Gibbs\cite{Tsal88,Kani01,Wang01}.

The idea of this work is to study the information or entropy from the outside of the
aforementioned formal systems based on the postulated information measures or
entropies. We just look at the geometry of the phase space of a non-equilibrium
system to calculate the information evolution and to see whether it has something to
do with the generalized entropies. We suppose an ensemble of non-equilibrium systems
moving in a fractal $\Gamma$-space, the volume of its initial condition gradually
mapping into fractal structure. If one looks at the trajectory of a system of the
ensemble as it runs over the permitted phase points, one will see that the phase
volume is covered more and more and that a fractal emerges from the covered regions.
In this case, as the system evolves in time, a scale refinement would be necessary
to calculate the heterogeneously occupied volume. So the long time effect of the
system of interest can be likened to its ensemble effect : the time behavior of the
trajectory of a system is replaced by the iteration of the scale refinement from the
initial phase volume of the ensemble. As a consequence, the information evolution in
time can be estimated by the ensemble average of the information growth during the
scale refinement, i.e.,
\begin{eqnarray}                                            \label{1}
{\cal I}(T) &=& \lim_{T\to\infty}\frac{1}{T}\int_0^T i(t)dt \Longrightarrow
\lim_{v_T\to\infty}\sum_{i=1}^{v_T}p_i\int_{s(i)}I_ids
-\sum_{j=1}^{v_0}p_j\int_{s(j)}I_jds  \\ \nonumber
&=&\lim_{v_k\to\infty}\sum_{i=1}^{v_k}p_i\int_{s(i)}I_ids
-\sum_{j=1}^{v_0}p_j\int_{s(j)}I_jds
\end{eqnarray}
where ${\cal I}(T)$ is the average information change during the time $T$, $i(t)$ is
the information change per unity of time, $p_i$ is the probability that the system
is found on the element $i$ of the phase space, $I_i$ is the density of information
on the element $i$ of volume $s(i)$, $v_T$ and $v_0$ are the total numbers of the
elements of phase space accessible to the system and visited by the trajectories at
time $t=T$ and $t=0$, respectively. According to our assumption, the $v_T$ elements
visited by the system form a fractal structure which can be reproduced by the $v_k$
elements yielded from the $v_0$ elements by certain map (scale refinement) of $k$
iterations. So we can put $v_k=v_T$.

Due to the incompleteness of the counting of state points or the calculation of
geometrical elements at any given scale in fractal structures\cite{Wang04}, the
discussion will be made on the basis of the normalization of incomplete probability
distribution\cite{Reny66} proposed for the complex systems having physical states
which are accessible to the systems but inaccessible to mathematical
treatment\cite{Wang01,Wang04,Wang02a,Wang03b}.

\section{Incomplete normalization}
Now we look at a fractal phase space of dimension $d_f$ embedded in a smooth
Euclidean space of dimension $d$. We know that any counting or calculation of state
number must be carried out at certain scale of the phase space. In fractal phase
space, the state number and the phase volume change from scale to scale. So the
state counting is never complete for a given scale or partition of the phase space.
In other words, the calculated states or phase volume is incomplete. Our method
consists in taking this incompleteness into account in the probability calculation.
We have assumed\cite{Wang01,Wang04,Wang02a,Wang03b} :
\begin{equation}                                            \label{2}
\sum_{i_k=1}^{v_k}p_{i_k}^q=1,
\end{equation}
where $v_k$ is only the number of the states or phase elements accessible to the
summation at the $k^{th}$ iteration, $q$ is given by $q=d_f/d$,
$p_{i_k}=s_{i_k}/S_0$ is the probability that the system visit the elements $i$ of
volume $s_{i_k}$ of the fractal at the $k^{th}$ iteration, and $S_0$ the volume of
the phase space containing the fractal.

The probability defined above is different from the usual frequency or time
definition. Here the probability $p_{i_k}$ does not sum to one because it is the
ratio of non-differentiable fractal elements to an integrable and differentiable
smooth space volume. So this definition allows one to carry out calculations of
fractal or hierarchical probability distributions by using the usual mathematical
tricks defined for smooth Euclidean space. It is analogous to the proposition in
\cite{Hilb94} to define probability $p_i$ for the system to visit the phase element
$i$ by the ratio of the number of trajectories ($\propto$ volume $s_i$) on the
element $i$ to the total number of trajectories ($\propto$ total volume $S_0$) in
the initial conditions uniformly distributed in a Euclidean space.

Eq.(\ref{2}) has been called incomplete normalization\cite{Wang01,Wang02a}. Its
incompleteness lies in the fact that the sum over all the $v_k$ elements at the
$k^{th}$ iteration does not mean the sum over all the possible states of the system
under consideration. In other words, the volume $s_{i_k}$ does not represent the
real number of states or trajectories on the element $i_k$ which, as expected for
any fractal and hierarchical structure, evolves with iteration or phase space
partition.

\section{Information growth due to fractal geometry}
The evolution of the accessible phase volume of a system during the scale refinement
is calculated as follows. The extra state points
\begin{equation}                                            \label{3}
\Delta_{i_k}=\sum_{j_{k+1}=1}^{n_{i_k}}s_{j_{k+1}}-s_{i_k}
\end{equation}
acquired from certain element $i_k$ at the iterate of $(k+1)^{th}$ order are just
the number of unaccessible states at $k^{th}$ order with respect to $(k+1)^{th}$
order, where $n_{i_k}$ is the number of elements $s_{j_{k+1}}$ replacing, at
$(k+1)^{th}$ iterate, the element $s_{i_k}$ and $j_{k+1}$ is the index of these
elements.

What is the information change in that case? At the iterate of order $k$, the
information content on $s_{i_k}$ is given by $I_k(i)=\int_{s_{i_k}} I(\rho)ds$ where
$I(\rho)$ is the information density as a function of the state density $\rho$
supposed scale invariant. This scale invariance, according to our assumption in
Eq.(\ref{1}), implies constant $\rho$ in time. So at $k+1$ order, we have
$I_{k+1}(i)=\sum_{j_{k+1}=1}^{n_{i_k}}\int_{s_{j_{k+1}}} I(\rho)ds$.

Remember that the definition of the probability $p_{i_k}\propto s_{i_k}$ implies
{\it constant $\rho$} over all the occupied elements of the phase space. In the case
where every phase point is visited with equal probability, constant $\rho$ means
constant state density over these elements. This is a natural result of the
uniformly distributed states in the initial condition phase volume $S_0$ if we
consider the time and scale invariance of $\rho$ mentioned above.

Then the information growth on certain phase element $i_k$ from $k^{th}$ to
$(k+1)^{th}$ iteration reads
\begin{equation}                                            \label{6}
\Delta I_k(i)=I_{k+1}(i)-I_k(i)=\int_{\Delta_{i_k}} I(\rho)ds =I(\rho)\Delta_{i_k}
\end{equation}
The relative information growth is then given by
\begin{eqnarray}                                            \label{7}
R_{k\rightarrow (k+1)}(i_k) = \Delta I_k(i)/I_k(i)
=\sum_{j_{k+1}=1}^{n_{i_k}}\frac{p_{j_{k+1}}}{p_{i_k}}-1.
\end{eqnarray}
The expectation of this relative information growth over all the fractal can be
calculated by using the unnormalized expectation as follows
\begin{eqnarray}                                            \label{7a}
\bar{R}_{k\rightarrow (k+1)} =\sum_{i_k=1}^{v_k}p_{i_k}R_{k\rightarrow (k+1)}(i_k)=
\sum_{i_{k+1}=1}^{v_{k+1}}p_{i_{k+1}}-\sum_{i_k=1}^{v_k}p_{i_k}.
\end{eqnarray}
The total relative information growth from $0^{th}$ to certain order, say,
$\lambda$, of the iteration is then given by
\begin{eqnarray}                                            \label{8}
R_\lambda=\bar{R}_{0\to\lambda} =\sum_{k=0}^{\lambda-1}\bar{R}_{k\rightarrow (k+1)}=
\sum_{i_\lambda=1}^{v_\lambda}p_{i_\lambda}-1 =
\sum_{i_\lambda=1}^{v_\lambda}(p_{i_\lambda}-p_{i_\lambda}^q)
\end{eqnarray}
since $\sum_{i_0=1}^{v_0}p_{i_0}=S_0/S_0=1$. We would like to mention that these
calculations can also be formally carried out under the formalism of complete
probability distribution if we suppose $\wp_i=p_i^q$. In this case, $R_k$ to the
$k^{th}$ iteration reads $R_k = \sum_{i_k=1}^{v_k}\wp_{i_k}^{1/q}-1$ with
$\sum_{i_k=1}^{v_k}\wp_{i_k}=1$. $\wp_{i_k}$ can be of course regarded as a
probability distribution on a complete ensemble of $v_k$ states.

Following are some properties of $R_\lambda$ (the index $\lambda$ will be dropped
from now on, i.e. $R=R_\lambda$) :

\begin{enumerate}
\item Nonadditivity : for a fractal of dimension $d_f$ composed of two sub-fractals
$A$ and $B$ of dimension $d_{f_A}$ and $d_{f_B}$ satisfying product joint
probability $p_{{i_A}{i_B}}=p_{i_A}p_{i_B}$, it is easy to show the following
nonadditivity :
\begin{eqnarray}                                            \label{11}
R(A+B)=R(A)+R(B)+R(A)R(B).
\end{eqnarray}

\item $R$ is positive and concave for $q>1$, and negative and convex for $q<1$. If
$q=1$ or $d=d_f$, the fractal structure does not exist any more, so $R=0$.

\item $R$ is an information growth attributed to the dimension difference $(d_f-d)$
and calculated from the actual probability distribution $p_i$. An interesting
feature of $R$ is that the ratio $\frac{R}{d_f-d}$ leads to the
Havrda-Charvat-Tsallis entropy $S=\frac{\sum_ip_i-\sum_i
p_i^q}{q-1}$\cite{Havrda,Tsal88}. The asymptote of this ratio for ${d_f\to d}$ leads
to Gibbs-Shannon entropy $S=-\sum_ip_i\ln p_i$. other nonadditive entropies in the
long list given by \cite{Esteban} can be obtained in similar way. For example, the
ratio $\frac{\ln(R+1)}{d_f-d}$ gives R\'enyi entropy $S^r=\frac{\ln\sum_i
\wp_i^{\alpha}}{1-\alpha}$ ($\alpha=1/q$) for complete distribution\cite{Reny66} or
$S^r=\frac{\ln\sum_i p_i}{q-1}$ for incomplete distribution\cite{Bashkirov}, and the
ratio $\frac{(R+1)^{d_f/d}-1}{d_f-d}$ gives Arimoto entropy
$S^a=\frac{(\sum_i\wp_i^{1/q})^q-1}{q-1}$\cite{Arimoto}.

\item It is straightforward to prove from the connections of $R$ with different
entropies that the maximum of these entropies is (mathematically) equivalent to the
extremization of the information growth $R$. So it is expected that the $R$-extremum
can be used to obtain probability distributions similar to those obtained from
maximum entropy. Here are some examples of the power law probability distributions
yielded by the extremization of $R$ ($q\neq 1$) for some chaotic systems.

The extremization of $R=\int_0^1\rho(x)dx-1$ ($0<x<1$ is the random variable of the
{\it continued fraction map}\cite{Beck} $x_{n+1}=1/x_n-\lfloor 1/x_n\rceil$) under
the constraints associated with the normalization $\int_0^1\rho^q(x)dx=1$ and the
unnormalized expectation $\overline{x}=\int_{0}^1\rho(x)xdx$ gives
\begin{eqnarray}                                            \label{12}
\rho(x)=\frac{1}{Z}\frac{1}{(1-\gamma x)^{1/(1-q)}}
\end{eqnarray}
where $q=1/2$ for the continued fraction map, $\gamma=-1$ is the Lagrange multiplier
associated with $\overline{x}$ and the constant $Z=\sqrt{\ln 2}$ is determined by
the incomplete normalization. The {\it Zipf-Mandelbrot's law}
$\rho(x)=\frac{A}{(1-\gamma x)^\alpha}$\cite{Alm} can be derived with
$1/q=1+1/\alpha$ ($A$ is the normalization constant). The distribution of the {\it
Ulam maps} $\rho(x)=\frac{1}{\pi(1-x^2)^{1/2}}$ ($-1<x<1$) can be obtained with
$q=1/3$ if $\overline{x^2}$ is used as a constraint.

\end{enumerate}

It should be emphasized that, although the same kind of distribution functions as
mentioned above can also be derived from the maximization of Tsallis entropy $S$ or
of R\'enyi one $S^r$ under the same constraints, the extremization of $R$ are
physically different from the entropy maximisation derived from the second law of
thermodynamics. $R$ is intrinsically connected with non-equilibrium evolving system.
So extremizing $R$ implies looking for the probability distribution that, among many
other possible ones, maximizes the information change of the system in evolution.
For this kind of systems, it is impossible to talk about maximizing entropy in the
sense of the second law because entropy is still in constant variation.

On the other hand, if the entropy of the system of interest is defined such that the
$R$-extremization is equivalent to the maximization of the entropy, then the
thermostatistics based on the maximum entropy principle may be discussed in
connection with the $R$-extremization. This is what we are doing in the following
for the statistical thermodynamics derived from maximizing Tsallis entropy.

\section{A thermodynamics of non-equilibrium systems with different $q$'s}
An important question about the nonextensive statistical thermodynamics based on
Tsallis entropy\cite{Tsal88} concerns its validity for the systems having different
$q$'s for which the thermodynamics must be formulated in a more general way than the
thermodynamics for the same $q$-systems\cite{Nauenberg,Tsal03}. This formulation is
crucial for nonextensive statistics because a composite system containing different
$q$-systems is a general case in nature. We are showing here that a possible
formulation can be made for {\it systems having different $q\neq 1$} on the basis of
the nonadditivity given by Eq.(\ref{11}). We suppose that the non-equilibrium system
is at some stationary state which maximizes Tsallis entropy for the total system
$A+B$. As mentioned above, this maximization is equivalent to the extremization of
$R$ for $A+B$, i.e., $dR(A+B)=0$. This leads to
$\frac{dR(A)}{1+R(A)}+\frac{dR(B)}{1+R(B)}=0$ which means
\begin{eqnarray}                                    \label{13}
\frac{(q_{A}-1)dS(A)}{\sum_ip_i(A)}+\frac{(q_{B}-1)dS(B)}{\sum_ip_i(B)}=0
\end{eqnarray}
where $S$ is the aforementioned Tsallis entropy. Now using the product joint
probability and the relationship $\sum_ip_i=Z^{q-1}+(q-1)\beta U$ connected with the
energy distribution function $p_i=\frac{1}{Z}[1-(q-1)\beta E_i]^{1/(q-1)}$ given by
the maximization of $S$, where $E_i$ is the energy of the system at state $i$ and
$U=\sum_ip_iE_i$ is the internal energy, we get
$\frac{(q_{A}-1)\beta(A)dU(A)}{\sum_ip_i(A)}
+\frac{(q_{B}-1)\beta(B)dU(B)}{\sum_ip_i(B)}=0$ which suggests following energy
nonadditivity
\begin{eqnarray}                                    \label{14}
\frac{(q_{A}-1)dU(A)}{\sum_ip_i(A)}+\frac{(q_{B}-1)dU(B)}{\sum_ip_i(B)}=0.
\end{eqnarray}
This relationship should be considered as the analog of the additive energy rule
$dU(A)+dU(B)=0$ of Boltzmann-Gibbs statistical thermodynamics. Eq.(\ref{13}) and
Eq.(\ref{14}) lead to
\begin{eqnarray}                                    \label{15}
\beta(A)=\beta(B)
\end{eqnarray}
where $\beta=\frac{\partial S}{\partial U}$ is the inverse temperature.

For the definition of pressure, as discussed in \cite{Wang03c}, Eq.(\ref{13})
finally leads to, at stationarity,
\begin{eqnarray}                                    \label{16}
\beta\left[P(A)\frac{(q_{A}-1)dV(A)}{\sum_{i}p_i(A)}
+P(B)\frac{(q_{B}-1)dV(B)}{\sum_{i}p_i(B)}\right]=0
\end{eqnarray}
where $P=\left(\frac{\partial U}{\partial V}\right)_S$ is the pressure and $V$ is
the volume. The intensive pressure, i.e. $P(A)=P(B)$, implies
\begin{eqnarray}                                    \label{17}
\frac{(q_{A}-1)dV(A)}{\sum_{i}p_i(A)} +\frac{(q_{B}-1)dV(B)}{\sum_{i}p_i(B)}=0
\end{eqnarray}
This volume can be interpreted as an effective volume allowing one to write the
first law as $dU=TdS-PdV$ in the case where the interface/surface effects cannot be
neglected compared to the volume effect. Un example of this kind of systems with
nonadditive effective volume is discussed in \cite{Wang03c}.

\section{Thermodynamics of nonadditive blackbody}
Now let us suppose a nonadditive blackbody obeying the above statistical
thermodynamics with the volume nonadditivity indicated by Eq.(\ref{17}). As
mentioned above, this case is possible when emission body is small (for example, the
thermal emission of nanoparticles or of small optical cavity) such that the
surface/interface effect may be important. We have seen in the above paragraph that
$dU$, $dS$ and $dV$ should be proportional to each other. This can be satisfied by
$U=f(T)V$ and $S=g(T)V$. In addition, we admit the photon pressure given by
$P=\frac{U}{3V}=\frac{1}{3}f(T)$. From the first law $dU=TdS-PdV$, we obtain
\begin{equation}                                        \label{18}
V\frac{\partial f}{\partial T}dT+fdV=T(V\frac{\partial g}{\partial
T}dT+gdV)-\frac{1}{3}fdV,
\end{equation}
which means $\frac{\partial f}{\partial T}=T\frac{\partial g}{\partial T}$ and
$\frac{4}{3}f=Tg$ leading to $\frac{1}{3}\frac{\partial f}{\partial
T}=\frac{4f}{3T}$. We finally get
\begin{equation}                                        \label{19}
f(T)=cT^4
\end{equation}
where $c$ is a constant. This is the Stefan-Boltzmann law. On the other hand, from
the relationship $(\frac{\partial S}{\partial V})_T=(\frac{\partial P}{\partial
T})_V$, we obtain $g=\frac{1}{3}\frac{\partial f}{\partial T}$ and $g(T)=bT^3$ where
$b$ is a constant. Notice that the above calculation is similar to that in the
conventional thermodynamics. This is because the thermodynamic functions here,
though nonadditive, are nevertheless ``extensive'' with respect to the effective
volume. This result contradicts what has been claimed for blackbody radiation on the
basis of non intensive pressure\cite{Nauenberg}, and is valid as far as the pressure
is intensive.

Following analysis shows what happens if one supposes additive volume $V$, i.e.
$dV(A)+dV(B)=0$. From Eq.(\ref{16}), one gets
\begin{eqnarray}                                    \label{20}
P(A)\frac{(q_{A}-1)}{\sum_{i}p_i(A)}=P(B)\frac{(q_{B}-1)}{\sum_{i}p_i(B)}.
\end{eqnarray}
So the conventional pressure $P=\left(\frac{\partial U}{\partial V}\right)_S$
becomes non intensive. If one still supposes $P=\frac{U}{3V}$, this will lead to a
deviation from the conventional Stefan-Boltzmann law as shown in \cite{Nauenberg}.
There are two questions here to be noticed. 1) Is non-intensive pressure possible?
2) The relationship $P=\frac{U}{3V}$ was established within the conventional
thermodynamics for additive photon gas. Is it still true for nonextensive or
nonadditive photon gas having non intensive pressure? Obviously, the final answers
to these questions require experimental proofs (which are still missing as far as we
know).

\section{Conclusion}
In summarizing, we have studied the information growth during long time evolution of
chaotic systems having fractal phase space through the scale refinement in the phase
space. This information growth turns out to take the trace form
$\sum_ip_i-\sum_ip_i^q$ which can be connected with several entropies generalizing
Shannon one. It is shown that the power law probability distributions of several
chaotic systems can be obtained by extremizing this information growth. However,
this work leaves open the questions as to whether in general one can maximize
entropy to get the probability distributions for non-equilibrium systems in
(stationary or not) evolution and whether one should extremize the entropy or
information change. In any case, if the information-entropy of the system is of
Tsallis type, these two methods turn out to be mathematically equivalent.

On the basis of the information growth, we discussed the thermodynamics of
non-equilibrium systems in stationary state which maximizes Tsallis entropy.
Intensive variables like temperature and pressure can be defined for an ensemble of
systems having different $q$'s. It is argued that Stefan-Boltzmann law for blackbody
radiation can be preserved within this thermodynamics. We would like to emphasize
that this work is carried out by using incomplete probability distribution and the
corresponding unnormalized expectation. We have noticed that the normalized
$q$-expectation $U=\sum_ip_i^qE_i$ for incomplete distribution could not be used due
to the product joint probability $p_{ij}(A+B)=p_i(A)p_i(B)$ connected with the
nonadditivity given by Eq.(\ref{11}) and Eq.(\ref{13}). In this case, the
unnormalized expectation allows one to split the thermodynamics of the composite
system into those of the subsystems, a necessary condition for the establishment of
zeroth law. This constraint on the nonextensive thermodynamics favours the use of
the unnormalized expectations $U=\sum_ip_iE_i$ at least for the systems having
different $q$'s. However, if the complete probability distribution $\sum_ip_i=1$ is
employed in the framework of the thermodynamics based on Tsallis entropy $S$, we
have $p_{ij}^{q_{A+B}}(A+B)=p_i^{q_A}(A)p_i^{q_B}(B)$ as generalized product joint
probability derived from Eq.(\ref{11}). In this case, the unnormalized expectation
$U=\sum_ip_i^qE_i$ or its normalized variation $U=\sum_ip_i^qE_i/\sum_ip_i^q$ should
be used in order to split the thermodynamics and to establish the zeroth law as
discussed in \cite{Wang03e}.

\end{document}